\documentclass[12pt]{article} 
\usepackage{graphicx}
\usepackage{caption}
\usepackage{subcaption}
\usepackage{pdflscape}
\usepackage[affil-it]{authblk}
\usepackage{amsmath}
\usepackage{amssymb}
\usepackage{color}

\begin{document}

\title{Phase field modelling of interfacial anisotropy driven faceting of precipitates}

\author{Arijit Roy}

\affil{Department of Metallurgical Engineering and Materials Science, Indian Institute of Technology Bombay, Powai, Mumbai 400076 INDIA.}

\author{E. S. Nani\footnote{Currently at Institute of Applied Materials - Computational Materials Science, Karlsruhe Institute of Technology, Karlsruhe 76131, GERMANY}}

\affil{Department of Metallurgical Engineering and Materials Science, Indian Institute of Technology Bombay, Powai, Mumbai 400076 INDIA.}

\author{Arka Lahiri}

\affil{Department of Materials Engineering, Indian Institute of Science, Bengaluru, 560012 INDIA.}

\author{M. P. Gururajan}

\affil{Department of Metallurgical Engineering and Materials Science, Indian Institute of Technology Bombay, Powai, Mumbai 400076 INDIA.\thanks{Corresponding author: $gururajan.mp@gmail.com$}}

\date{}

\maketitle

\section*{Abstract}
We use extended Cahn-Hilliard (ECH) equations to study faceted precipitate morphologies; specifically, we obtain four sided precipitates (in 2-D) and dodecahedron (in 3-D) in a system with cubic anisotropy, and, six-sided precipitates (in 2-D, in the basal plane), hexagonal dipyramids and hexagonal prisms (in 3-D) in systems with hexagonal anisotropy. Our listing of these ECH equations is fairly comprehensive and complete (upto sixth rank tensor terms of the Taylor expansion of the free energy). We also show how the parameters that enter the model are to be obtained if either
the interfacial energy anisotropy or the equilibrium morphology of the precipitate is known.

\section{Introduction}

Properties of crystalline materials are anisotropic due to the anistropy of the underlying continuum. In particular, the interfacial 
energy in crystalline systems is anisotropic and can have a strong influence on the formation and evolution of microstructures. Phase field 
models, which are best suited for the study of the formation and evolution of microstructures 
(see~\cite{ChenAnnRevMatSci,BoettingerEtAlAnnRevMatSci, Steinbach,Voorhees,LeuvenGang} for some recent reviews), have been used 
quite successfully to study the effect of interfacial energy anisotropy on microstructures and their evolution: see
~\cite{AbiHaider,Lowengrub,NaniGururajan,Haxhimali2006,QinBhadeshia1,QinBhadeshia2,Ian,ChoiEtAl2015,WangEtAl,MoelansEtAl,
Wheeler,McFaddenEtAl,Langer,BraunEtAl,BraunEtAl2,CahnEtAl,Eggleston2001,HeoSaswataChen,JinWangKhachaturyan,
WangBanerjeeSuKhachaturyan,WangJinCuitinoKhachaturyan,WangJinKhachaturyan,ZhangChen,Sekerka,Loginova2004}
for some representative examples. 

The phase field models that incorporate interfacial energy anisotropy do so in one of two ways: (A) replace the gradient energy coefficient by an anisotropic function or polynomial, and (B) include higher order terms in the Taylor series expansion of the free energy functional. In this paper, we take the second approach -- which is an extension of the original Cahn-Hilliard equation along the lines shown by Abinandanan and Haider~\cite{AbiHaider} (hereafter referred to as ECHAH) and Torabi and Lowengrub~\cite{Lowengrub} (hereafter referred to as ECHTL); 
this approach is argued to be advantageous in terms of the levels of anisotropy one can incorporate~\cite{AbiHaider} and the kinetics remaining diffusion-limited~\cite{ChoiEtAl2015}; in addition, the first method requires regularisation~\cite{Eggleston2001} for large anisotropies while ours does not. 

During solid-solid phase transformations interfacial energy anisotropy is known to lead to faceted precipitates: 
see for example, PbS precipitates in Na-doped PbTe system~\cite{HeEtAl2012}, Al$_3$Sc precipitates in Al(Sc) alloys~\cite{MarquisSeidman2001}, Pt precipitates in sapphire~\cite{SantalaEtAl}, several metallic precipitates in internally reduced oxides~\cite{BackhausRicoult1997}, and  Al$_3$Ti precipitates in Al~\cite{vandeWalleEtAl2014}. Our objective in this paper is to obtain, using phase field modelling, faceted precipitates in 
cubic systems that distinguish between $\langle 100 \rangle$ and $\langle 111 \rangle$ (for which one has to necessarily include fourth rank tensor terms~\cite{AbiHaider}) and those that prefer $\langle 110 \rangle$ over both $\langle 100 \rangle$ and $\langle 111 \rangle$ (for which 
one has to necessarily include sixth rank tensor terms~\cite{NaniGururajan}); additionally,  the sixth rank tensor terms can also be used to
study systems with hexagonal anisotropy~\cite{NaniGururajan}.  

In this paper, after a brief description of the model and its numerical implementation (Sections~\ref{Section2} and~\ref{Section3}), 
we present 1-, 2- and 3-D results; specifically, we show a variety of faceted precipitates in 3-D obtained using our model (Section~\ref{Section4})
and conclude the paper with a summary of the salient results from this study (Section~\ref{Section5}). 

\section{Formulation} \label{Section2}

We consider a binary alloy system. We assume that its microstructure is completely described by the 
(coarse-grained) local composition ($c$), and the local gradients $c_i$, curvature $c_{ij}$ and 
aberration of the composition $c_{ijk}$ --  that is, $c_i = \frac{\partial c}{\partial x_i}$, 
$c_{ij} = \frac{\partial^2 c}{\partial x_i \partial x_j}$, and, 
$c_{ijk} = \frac{\partial^3 c}{\partial x_i \partial x_j \partial x_k}$ where $x_i$ is the position vector. 
We can write the free energy functional of such a system as follows:
\begin{equation} \label{FreeEnergyFunctional}
F = N_V \int f dV
\end{equation}
where, $N_V$ is the number density of atoms or molecules in the system, $V$ is the volume,
$F$ is the free energy and $f$ is the free energy per atom. 
Assuming that the underlying crystalline continuum is centro-symmetric, in order
to incorporate cubic and hexagonal anisotropies,
$f$ can be written as follows (see~\cite{NaniGururajan} for details):
\begin{eqnarray} \label{FinalFreeEnergy}
f & = &  f_0 + P(c_1,c_2,c_3) + Q(c_{11},c_{22},c_{33},c_{12},c_{23},c_{13}) \\  \nonumber
& & + R(c_{111},c_{222},c_{333},c_{112},c_{113},c_{221},c_{223},c_{331},c_{332},c_{123})
\end{eqnarray}
where $f_0$ is the bulk free energy density (assumed to be a double well potential, namely, $A c^2 (1-c)^2$), 
and, $P$, $Q$ and $R$ are homogeneous polynomials.  
In Table~\ref{Table1}, we list the forms of these polynomials: 
$P$ consists of three parts: homogeneous polynomials of orders 2 (denoted by P$^2$), 4 (denoted by P$^4$) and 
6 (denoted by P$^6$); $Q$ and $R$ are homogeneous polynomials of order 2. 
The coefficients of these polynomials are assumed to be constants.
\begin{table}[tbh] 
\caption{\label{Table1} Polynomials $P$, $Q$ and $R$.}
\begin{tabular}{|c|c|c|}
\hline \hline
{\bf Polynomial } & {\bf Order} & {\bf Form} \\
\hline \hline
$P^2$ & 2&$p_1 (c_1^2 + c_2^2) + p_2 c_3^2$ \\
\hline
$P^4$& 4 & $m_1 (c_1^4+c_2^4) + 2 m_2 c_1^2 c_2^2 + m_3 c_3^4 + 2 m_4 (c_1^2 + c_2^2) c_3^2$ \\
\hline
$P^6$& 6 & $n_1 (c_1^2+c_2^2)^3 + n_2 c_3^6$ \\ 
&&+$n_3 c_1^2(c_1^2-3c_2^2)^2 + n_4(c_1^2 +c_2^2)^2 c_3^2 + n_5 (c_1^2+c_2^2) c_3^4$ \\
&&+$n_6 (c_1^2+c_2^2+c_3^2)(c_1^2 c_2^2 +c_2^2 c_3^2 + c_3^2 c_1^2) + n_7 c_1^2 c_2^2 c_3^2$ \\
\hline \hline
$Q$ & 2 & $q_1 (c_{11}^2 +c_{22}^2) + q_2 c_{33}^2$ \\
 && $+ 2 q_{3} c_{11}c_{22} + 2 q_4 (c_{11} + c_{22}) c_{33}$  \\
 && $+q_5 (c_{13}^2 + c_{23}^2) +  q_6 c_{12}^2$ \\
\hline \hline
$R$ & 2 & $r_1 c_{111}^2 + r_2 c_{333}^2 + r_{3} c_{112}^2 + r_4 c_{221}^2$ \\
&& $+ 2 r_5 c_{113} c_{223} + 2 r_6 (c_{111}c_{331} + c_{222} c_{332}) + 2 r_7 (c_{112} c_{332} + c_{221}  c_{331})$ \\
&& $+ r_8 (c_{113}^2+ c_{223}^2) + r_9 (c_{331}^2 + c_{332}^2) + 2 r_{10} c_{333}(c_{113}$ \\
&&  $+ c_{223}) + r_{11} c_{222}^2 + r_{12}  c_{111}c_{221}$ \\
&& $+ r_{13}  c_{222}c_{112} + r_{14}  c_{123}^2$ \\
\hline \hline
\end{tabular}
\end{table}

For the polynomials listed in Table~\ref{Table1}, the choice of coefficients determines whether the 
interfacial free energy is isotropic or cubic or hexagonal anisotropic; hence,
in Table~\ref{Table2} we list the inter-relationships 
and the constraints on the coefficients of these polynomials for all these three cases (assuming term wise positive definiteness~\cite{NaniGururajan}).

\begin{landscape}
\begin{table}[ptbh]
\caption{\label{Table2} Constraints on the coefficients of the polynomials listed in Table~\ref{Table1}.}
\begin{tabular}{|c|c|c|c|}
\hline \hline
 & {\bf Isotropic} & {\bf Cubic} & {\bf Hexagonal}\\
\hline \hline
$P^2$ & $p_1 = p_2 \geq 0$ & $p_1 = p_2 \geq 0$ & $p_1 \geq 0$; $p_2 \geq 0$ \\
\hline
$P^4$ &$m_1 = m_2 = m_3 = m_4$ & $m_1 = m_3 \geq 0$; & $m_1 = m_2 \geq 0$; $m_3 \geq 0$;\\
 &$m_1 \geq 0$& $m_2 = m_4 \geq -m_1$;& $m_4 \geq - \sqrt{m_1 m_3}$\\
\hline
$P^6$  &$n_1 = n_2 = \frac{n_4}{3} = \frac{n_5}{3}$&$n_1 = n_2 = \frac{n_4}{3} = \frac{n_5}{3}$&$n_1 \geq 0$; $n_2 \geq 0$; $n_3 \geq -n_1$; $n_6 = n_7 = 0$;\\
& $n_1 \geq 0$&$n_3 = 0$ &Either $n_4,n_5 \geq 0$  \\
&$n_3 = n_6 = n_7 = 0$ & $n_6 \geq - 4 n_1$ &or they satisfy the equation\\
& &$n_7 \geq -9 (3 n_1 + n_6)$ & $4 X n_5^3 + 4 n_2 n_5^3 + 27 X^2 n_2^2$\\
&&&  $- 18X n_2 n_4 n_5 -n_4^2 n_5^2 \geq 0$\\ 
&&& where $X = \mathrm{min} \{ n_1, n_1+n_3\}$\\
\hline \hline
$Q$ & $q_1=q_2 \geq 0$; $q_3=q_4$;  &$q_1=q_2$; $q_3 = q_4$ & $q_1 \geq 0$; $q_2 \geq 0$; $-q_1 \leq q_3 \leq q_1$\\
 &$q_5=q_6 = 2(q_1-q_3)$ &$q_5 = q_6$&$q_4^2 \leq \frac{1}{2}(q_1+q_3)q_2$; $q_5 \geq 0$ \\
 &$-\frac{q_1}{2} \leq q_3 \leq q_1$  &$q_1$, $q_5 \geq 0$; $-\frac{q_1}{2} \leq q_3 \leq q_1$&$q_6 = 2 (q_1-q_3)$\\
\hline \hline
$R$ &$r_1=r_2 = r_{11} \geq 0$ & $r_1 = r_2  = r_{11}\geq 0$& $r_1$,$r_2$,$r_8$,$r_9 \geq 0$; $r_6 = r_7$; $r_1 \leq r_4 \leq 9 r_1$ \\
& $r_3=r_4=r_8=r_9$ &$r_3 = r_4 = r_8 = r_9 \geq 0$ &$r_3 \geq \frac{(9 r_1-r_4)}{8}$; $r_5^2 \leq \frac{r_9(9 r_1-r_4)}{8}$\\
&$r_5=r_6=r_7 = r_{10} = 3r_1-r_4$ & $r_6 = r_{10}$; $r_{12} = r_{13} = 2 r_6 $ &$r_{10}^2 \leq r_2 r_8$\\
& $r_1 \leq r_3 \leq 6 r_1$  & $r_6^2 \leq r_1 r_3$ &$\frac{2r_{10}^2}{r_2} - r_8 \leq r_5 \leq r_8$\\
&$r_{12} = r_{13} = 2 r_5$ &$r_5 = r_7$; $r_{14} \geq 0$ & $r_{11} = \left[r_1+\frac{1}{9}(r_3-r_4)\right]$\\
&$r_{14} = 3 (r_4-r_1)$ & $\frac{2 r_6^2}{r_1} - r_3 \leq r_5 \leq r_6$ & $r_{12} = (3r_1-r_4)$; $r_{14} = 3 (r_7-r_5)$ \\
&  &  & $r_{13} = \left( 3r_1-\frac{2}{3}r_3-\frac{1}{3}r_4 \right)$\\
\hline \hline
\end{tabular}
\end{table}
\end{landscape}

Once the free energy is given, the 
variational derivative of the free energy functional with composition gives the relevant
chemical potential ($\mu$)~\cite{AbiHaider,ShamesDym2013}:
\begin{equation}\label{ChemicalPotential}
\mu = \frac{\delta F}{\delta c} = \frac{\partial f_0}{\partial c} - \frac{\partial}{\partial x_i}  \left[ \frac{\partial F}{\partial c_i} \right]  + \frac{\partial^2}{\partial x_i \partial x_j} \left[ \frac{\partial F}{\partial c_{ij}} \right] - \frac{\partial^3}{\partial x_i \partial x_j \partial x_k} \left[ \frac{\partial F}{\partial c_{ijk}} \right]
\end{equation}
where we have used Einstein summation convention, namely, that repeated indices are summed.

Assuming the mobility $M$ to be a constant and accounting for the constraint that the composition is a conserved order parameter, we obtain
the Cahn-Hilliard equation as follows: 
\begin{equation}
\label{E:CH_eq}
\frac{\partial c}{\partial t} = M \nabla^2 \left[ {\mu}_0 - \mu_{c_i} + \mu_{c_{ij}} - \mu_{c_{ijk}} \right]
\end{equation}
where the $\mu_0 = \frac{\partial f_0}{\partial c}$, $\mu_{c_i} = \frac{\partial}{\partial x_i}  \left[ \frac{\partial F}{\partial c_i} \right] $, $\mu_{c_{ij}} = \frac{\partial^2}{\partial x_i \partial x_j} \left[ \frac{\partial F}{\partial c_{ij}} \right]$, and, $\mu_{c_{ijk}} = \frac{\partial^3}{\partial x_i \partial x_j \partial x_k} \left[ \frac{\partial F}{\partial c_{ijk}} \right]$. In Table~\ref{Table3}, we list these chemical potential terms for each of the polynomials listed in Table~\ref{Table1}. From Table~\ref{Table3}, it is clear that the second order polynomials, be in gradient, or curvature or aberration, lead to linear terms in the evolution equation while the fourth and sixth order polynomials in gradients lead to highly nonlinear terms in the extended Cahn-Hilliard equation. Though these nonlinear terms seems to make the extended Cahn-Hilliard equation very stiff, we show below that they can be solved (albeit with time-steps that are smaller by a factor of 10$^4$) using Fourier spectral techniques. It is also possible to use finite difference technique to solve these highly non-linear equations -- for example, the Fig.~\ref{F:3D_grad_6_dodeca} is obtained using a finite difference implementation.

Here, we wish to emphasise that the polynomials in Table~\ref{Table1} describe a family of (extended) Cahn-Hilliard equations.
By appropriate choice, we obtain classical Cahn-Hilliard equation (only $P^2$ is non-zero with $p_1=p_2$),
ECHAH (only $P^{2}$ and $Q$ are non-zero) and ECHTL (only $P^2$, $P^4$ and $Q$ are non-zero) from this table.
To obtain hexagonal anisotropy, we have to incorporate either $P^{6}$ or $R$ or both (with or without any of the other terms, 
namely, $P^{2}$, $P^{4}$ and $Q$). It is also possible to obtain cubic anisotropy by incorporating $P^{4}$ or $Q$ or $P^{6}$ or $R$ or 
any combination of them with or without $P^{2}$. Specifically, in a cubic system, to make (110) planes preferred over both (100) and (111), 
we need to necessarily incorporate either $P^6$ or $R$ or both. However, the most important point to note is that the parameters for 
each of the tensors have to be chosen appropriately to obtain the required anisotropy -- as discussed in detail below in Sec.~\ref{ChoiceOfParameters}.

A note on term wise positive definiteness is due here; similar to the argument of ECHAH 
(below Eq.~35 in~\cite{AbiHaider}), from the chemical potential expressions listed in Table~\ref{Table3}, using
Fourier transform of the RHS of the evolution equation, term wise positive definiteness arguments can be made  
(at least in some cases). Such term wise positive definiteness allows us to pick and 
choose any combination of these terms as described in the previous paragraph. 

Note that our evolution equations are different from those used in the solidification studies -- by Qin and Bhadeshia~\cite{QinBhadeshia1,QinBhadeshia2} for cubic and hexagonal interfacial energy anisotropies and Haxhimali et al~\cite{Haxhimali2006} for orientation selection in dendrites in cubic systems in which the preference for dendritic arms changes from $\langle 100 \rangle$ to $\langle 110 \rangle$. Both Qin and Bhadeshia and Haxhimali et al  assume that the gradient energy coefficient $\kappa$ is a function of $n_x$, $n_y$ and $n_z$. On the other hand, we have used higher order terms in the Taylor series expansion of the free energy. In addition, the polynomial terms used by Haxhimali et al~\cite{Haxhimali2006} would correspond to gradient sixth rank tensors (in our formulation) while those used by Qin and Bhadeshia would correspond to gradient eighth and six rank tensors for cubic and hexagonal systems respectively. 

Finally, in many solid-solid phase transformations, interface coherency and hence elastic energy effects can be very important. Our model 
can be extended to include elastic anisotropy induced faceting and studies on the competing effects of interfacial and elastic anisotropy on precipitate morphologies is in progress.

\begin{landscape}
\begin{table}[tbhp] 
\caption{\label{Table3} The chemical potentials $\mu_{c_i}$, $\mu_{c_{ij}}$ and $\mu_{c_{ijk}}$.}
\begin{tabular}{|c|c|}
\hline \hline
 ${\mathbf{\mu}}$  & {\bf Expression} \\
\hline \hline
$\mu_{c_i}(2)$ &$p_1 (c_{11} + c_{22}) + p_2 c_{33}$ \\
\hline
$\mu_{c_i}(4)$ & $12 m_1 c_1^2 c_{11} + 12 m_1 c_2^2 c_{22} + 12 m_3 c_3^2 c_{33} + 16 m_2 c_1 c_2 c_{12} + 16 m_4 c_3 (c_1 c_{31} +  c_2  c_{23})$ \\
 &  $4 m_2 c_{11} c_2^2 + 4 m_4 c_{11} c_3^2 + 4 m_2 c_{22} c_1^2 + 4 m_4 c_{22} c_3^2 + 4 m_4 c_{33} (c_1^2 + c_2^2)$ \\
\hline
$\mu_{c_i}(6)$ & $ 6 n_1 (c_{11}+c_{22}) (c_1^2+c_2^2)^2 + 24 n_1 (c_1^2+c_2^2)(c_1 c_1 c_{11} + 2 c_2 c_1 c_{21} + c_2 c_2 c_{22}) + 30 n_2 c_3^4 c_{33}$\\
& $+ 2 n_3 c_{11} (c_1^2-3c_2^2)^2 + 8 n_3 c_1 (c_1^2-3c_2^2) (c_1 c_{11} -3 c_2 c_{21}) + 12 n_3 c_1^2 (c_{11}-c_{22}) (c_1^2-3c_2^2)$ \\
&  $+ 8 n_3 c_1^3 (c_1 c_{11} -3 c_2 c_{21})  - 24 n_3 c_1 c_2 c_{12} (c_1^2-3c_2^2) - 24 n_3 c_2 c_1^2 (c_1 c_{12} -3 c_2 c_{22})$ \\
& $+4 n_4 c_3^2 (c_1^2+c_2^2) (c_{11} + c_{22}) + 16 n_4 c_3 (c_1^2+c_2^2) (c_1 c_{31} + c_2 c_{32}) + 2 n_4 c_{33} (c_1^2+c_2^2)^2$\\
& $+ 8 n_4 c_3^2 (c_1 c_1 c_{11} + 2 c_2 c_1 c_{21} + c_2 c_2 c_{22})  + 2 n_5 (c_{11}+c_{22}) c_3^4  + 16 n_5 c_3^3 (c_1 c_{31} + c_2 c_{23}) $\\
& $+12 n_5 c_3^2 c_{33} (c_1^2+c_2^2) + 2 n_6 (c_{11} + c_{22} + c_{33}) (c_1^2 c_2^2 + c_2^2 c_3^2 + c_3^2 c_1^2))$\\
& $+8 n_6 c_1 c_2 c_{12} (c_1^2 + c_2^2 + 2c_3^2) +8 n_6 c_1 c_3 c_{31} (c_1^2 + 2 c_2^2 + c_3^2) $ \\
& $+8 n_6 c_2 c_3 c_{23} (2 c_1^2 + c_2^2 + c_3^2) + 4 n_6 [ c_1^2 c_2^2 (c_{11} + c_{22}) + c_1^2 c_3^2 (c_{11} + c_{33}) + c_2^2 c_3^2 (c_{22} + c_{33})] $ \\
& $2 n_6 (c_1^2+c_2^2 +c_3^2) [c_{11} (c_2^2+c_3^2 )+ c_{22} (c_1^2+c_3^2) + c_{33} (c_1^2+c_2^2)]$\\
& $ + 8 n_6 (c_1^2+c_2^2 +c_3^2) (c_1 c_2 c_{12} + c_1 c_3 c_{31} + c_2 c_3 c_{23})$\\
& $+4 n_6 [c_1^2 c_{11} (c_2^2+c_3^2) + c_2^2 c_{22} (c_1^2+c_3^2)+ c_3^2 c_{33} (c_1^2+c_2^2)] $ \\
& $2 n_7 (c_{11} c_2^2 c_3^2 + c_{22}  c_1^2 c_3^2  + c_{33} c_1^2 c_2^2) + 8 n_7 (c_1 c_2 c_{12} c_3^2 + c_1 c_3 c_{31} c_2^2 +c_2 c_3 c_{23} c_1^2) $ \\
\hline \hline
$\mu_{c_{ij}}$ &  $2 q_1 (c_{1111} +c_{2222}) + 2 q_2 c_{3333}$ $+ 4 q_{3} c_{1122} + 4 q_4 (c_{3311} + c_{2233}) +2 q_5 (c_{1313} + c_{2323}) +  2 q_6 c_{1212}$ \\
\hline \hline
$\mu_{c_{ijk}}$ &  $2 r_1 c_{111111} + r_2 c_{333333} + (2 r_{3} + 2 r_{12}) c_{111122} + (2 r_4 + 2 r_{13}) c_{222211} + (4 r_5 + 8 r_7 + 2 r_{14}) c_{112233}$\\
&  $+ (4 r_6 + 2 r_8) (c_{111133}+ c_{222233} ) + (2 r_9 + 4 r_{10}) (c_{333311} + c_{333322}) + 2 r_{11} c_{222222} $ \\
\hline \hline
\end{tabular}
\end{table}
\end{landscape}

\section{Numerical Implementation} \label{Section3}

We have used semi-implicit and explicit Fourier spectral implementation to solve the phase field equations~\ref{E:CH_eq} described in Sec.~\ref{Section2}. The chemical potential ($\mu$) expressions (listed in Table.~\ref{Table3}), as noted earlier, from the point of view of numerical implementation, are of two distinct types -- namely, those that are amenable to semi-implicit method and those are amenable only to the explicit method. The second rank gradient, curvature and aberration tensor terms lead to  a term which is linear in $c$ in the evolution equation and hence are implemented in the same way as they were implemented by ECHAH, namely, using a semi-implicit Fourier spectral technique~\cite{ChenShen1998}. The fourth and sixth rank gradient tensor terms, on the other hand,  lead to the non-linear terms in which the gradient and curvature terms are coupled; see Table.~\ref{Table3}. In such a case, the implementation is explicit (and is slightly more involved). For example, consider the following term on the right hand side of the evolution equation~\ref{E:CH_eq}: $\nabla^2 \left[ \mu_{c_i} (4) \right]$. The $\mu_{c_i}$ terms are of the type $c_{ij} c_k c_l$. Hence, in the numerical implementation, (i) first these terms (namely, $c_{ij}$ and $c_{k}$) are evaluated in Fourier space (since the derivatives are evaluated far more easily and accurately in the Fourier space); (ii) then, these curvature and gradient terms are brought to real space; and, (iii) the relevant $\mu_{c_i}$ term is calculated in the real space; and, (iv) finally, the non-linear term is taken to the Fourier space to be used in the evolution equation. So, in this case, there is an extra evaluation of forward and inverse Fourier transforms; and, the derivative terms are evaluated in the current time (and hence, explicitly). Thus, the implementation for these two cases is explicit. Hence, from a numerical implementation point of view, the curvature ($Q$) or aberration ($R$) based higher order tensor terms are preferred over gradient terms ($P^4$ and $P^6$).

The numerical implementation is carried out on the non-dimensionalised evolution equations. The non-dimensionalisation is the same as that described in~\cite{AbiHaider}: namely, $A$ in the bulk free energy density $f_0$ is used as the characteristic energy and the lattice parameter $a$ is used as the characteristic length. The characteristic time is chosen such that the mobility $M$ is unity. Note that the composition $c$ used in the bulk free energy density is scaled such that the equilibrium matrix and precipitate compositions ($c_e^m$ and $c_e^p$, respectively) are zero and unity, respectively. This non-dimensionalisation results in
the $A$ and $M$ parameters taking a value of unity. The far-field composition in the matrix is denoted by $c_\infty$ and is chosen to be 0.2 (in 2-D simulations) and 0.1 (in 3-D simulations). The grid spacing for spatial variables $\Delta x = \Delta y = \Delta z = 0.5 $ (in 1- and 2-D simulations) and $\Delta x = \Delta y = \Delta z = 1.0 $ (in 3-D simulations).
The time step used in these simulations are $\Delta t = 10^{-5}$ (for $P^4$ and $P^6$) and $\Delta t = 10^{-1}$ (for $R$). The 2-D simulations are carried out on a $512 \times 512$ grid while the 3-D simulations are carried out on $128 \times 128 \times 128$ grid. We list only the independent tensor coefficients (described in Tables.~\ref{Table1} and~\ref{Table2}) used in the simulations are described in Sec.~\ref{Section4} at the appropriate places; the dependent parameters (such as $r_{12}$ for example) are obtained using the relationships listed in Table.~\ref{Table2}.

\section{Results} \label{Section4}

In this section, we first present the results from our 1-D simulations. These results help us understand the composition profiles, 
the scaling behaviour of interfacial energy and interfacial width with the tensor coefficients, and in determining the Wulff plots. The 2- and 3-D simulation results are then presented to show the precipitate morphologies that are obtained.

\subsection{Composition profiles across planar interfaces}

\begin{figure}[h]
\centering
\includegraphics[height=2.2in,width=2.2in]{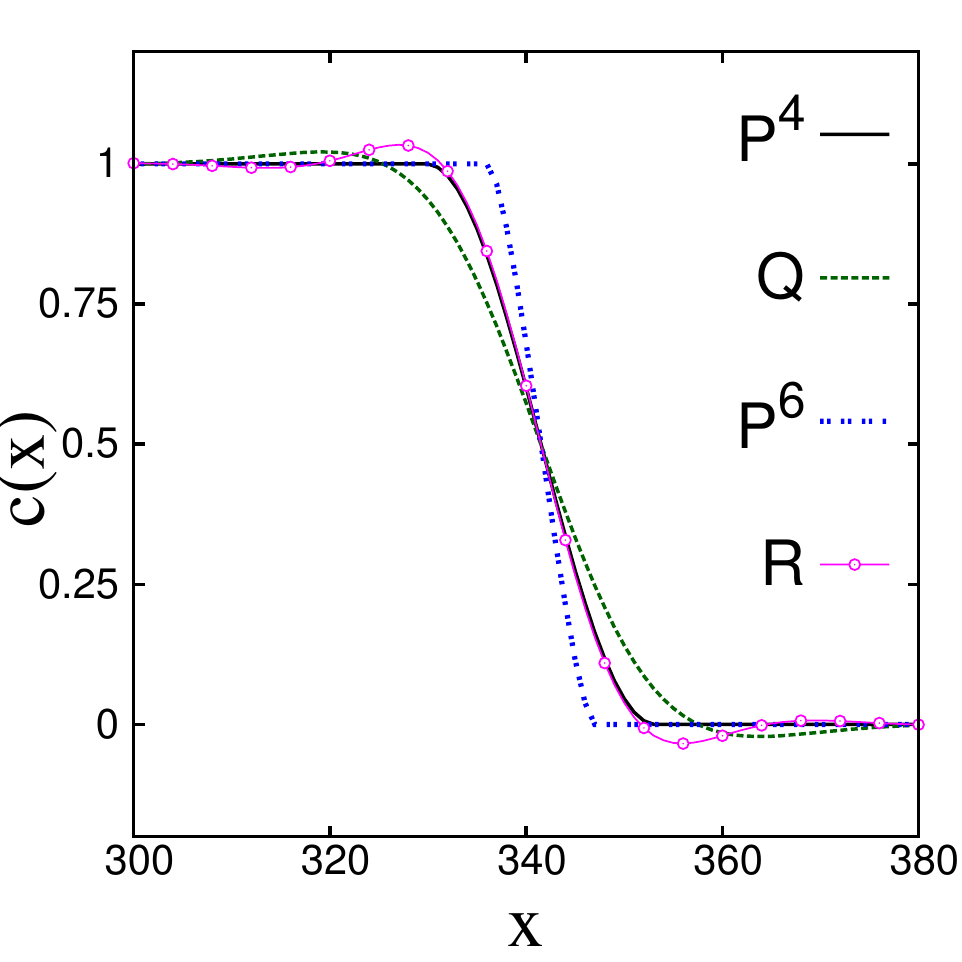}
\includegraphics[height=2.2in,width=2.2in]{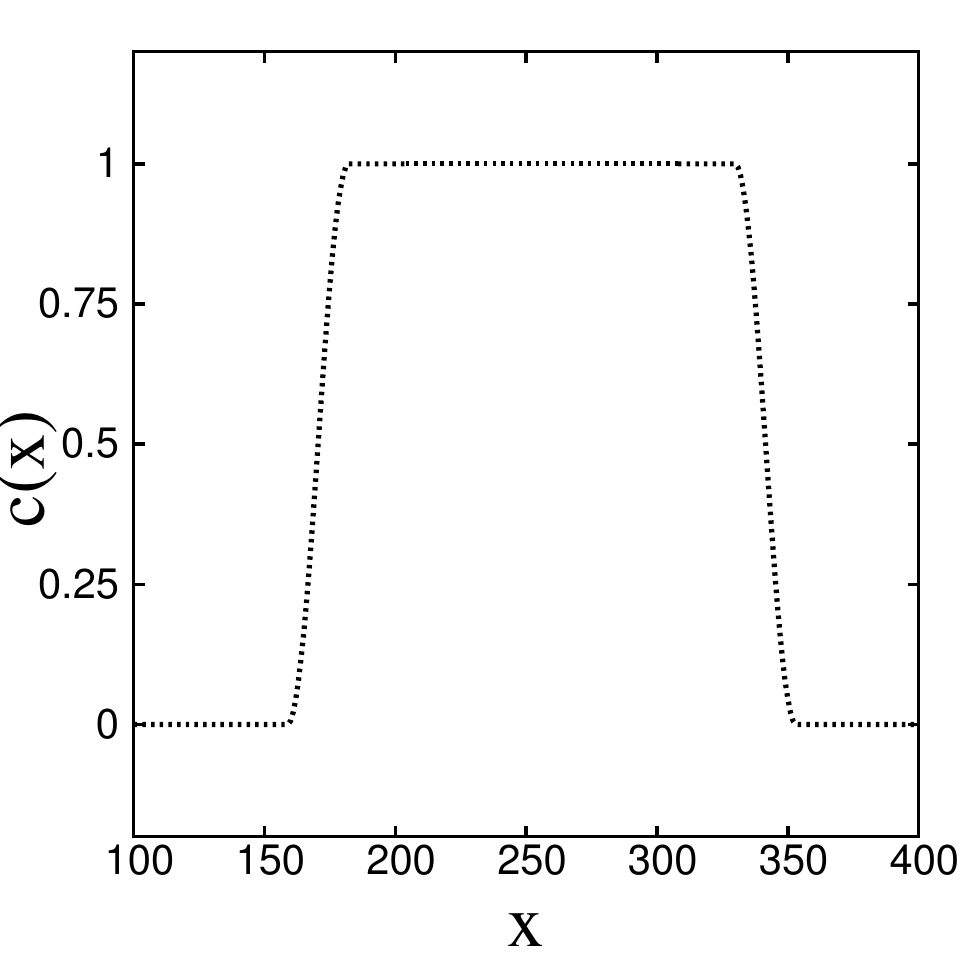}
\caption{Variation of composition $c(x)$ across a planer interface for different tensor coefficients (left) and an example of a complete two-interface composition profile with periodic boundary condition (right).}\label{F:1D_comp}
\end{figure}

In Fig.~\ref{F:1D_comp} we show the equilibrium composition profile across a planar interface in four different cases, namely, with $P^4$, $P^6$, $Q$ and $R$. In all these cases, except for the mentioned tensor coefficient, the rest of the coefficients were assumed to be identically zero. For example, in the $P^4$ case, all the other coefficients, namely, $P^2$, $P^6$, $Q$ and $R$ were assumed to be identically zero. Further, in all the cases, the interface normal was assumed to be along the x-direction of the simulation cell and the effective coefficient value is assumed to be numerically the same, namely, 64. To be consistent with the periodic boundary conditions, two planar interfaces are introduced in the simulation cell as shown in the right-side of Fig.~\ref{F:1D_comp}. The simulation was run for a long time till equilibrium of the composition profiles were achieved. For the sake of clarity, we zoom in on one of the equilibrated interfaces. 

As noted by ECHAH, the curvature terms lead to an interface profile in which to the left and right of the central point (that is, the point where c=0.5), the composition profile goes above and below unity and zero respectively. This is due to the effect of the curvature terms. In a similar manner, in the case of aberration, the composition profile goes above and then below unity (below and above zero) before becoming unity (zero) in the bulk. On the other hand, the gradient coefficients always lead to tanh-like profiles.  As noted above, from a semi-implicit Fourier spectral technique implementation point of view, it is preferred to choose curvature / aberration terms since they will use large time steps. However, such a choice has this unwanted side effect of producing ripples in the interface profile. If such ripples are not preferred, then gradient terms are to be chosen. Having said that, in terms of anisotropy, as we show below, all these terms behave in the same manner.

\subsection{Scaling of interfacial free energy and interfacial width}

From the 1-D profiles (like the ones shown above in Fig.~\ref{F:1D_comp}), it is possible to calculate both the interfacial energy and interfacial width~\cite{Cahn-Hilliard_1958}. For the assumed bulk free energy density, the total interfacial energy is but the free energy in Eq.~\ref{FreeEnergyFunctional} (assuming that the 1-D profiles are for a system with unit cross-sectional area). Since there are two interfaces
in our simulations, by dividing the total interfacial energy by two, we obtain the interfacial free energy ($\sigma$). The interfacial width ($w$)
is the distance between the intersection points of the straight line drawn at 0.5 composition with zero and unity respectively: 
\begin{equation}
 w = \Bigg(\Bigg|\frac{\partial c}{\partial x} \Bigg|_{0.5}\Bigg)^{-1} \label{E:width}
\end{equation}
Note that we use this definition even in the case where the curvature and aberration terms cause ripples in the bulk close to the interface.

Cahn and Hilliard~\cite{Cahn-Hilliard_1958} have shown that both the interfacial energy and interfacial width scale with the second rank gradient term ($P^2$ in our notation) as $(P^2)^{0.5}$. In the case of ECHAH~\cite{AbiHaider}, it was shown that the interfacial energy and interfacial width scale as $(Q)^{0.25}$ (for the fourth rank  term, $Q$ in our terminology). Using the same arguments, namely, that both the interfacial energy $\sigma$ and the interfacial width $w$ are homogeneous functions of order unity, we can show that the interfacial energy and widths are expected to scale as $(P^{4})^{0.25}$, $(P^6)^{0.167}$, and, $(R)^{0.167}$. In other words, given that $n$ is the order of any given coefficent tensor $\alpha$, the interfacial energy and width scale as $(\alpha)^{\frac{1}{n}}$. 

Let us define the scaling parameter $\chi$ as $(P^{2})^{0.5}$. The scaled interfacial energy $\sigma_s$ and interface width $w_s$ are defined as $\sigma/\chi$ and $w/\chi$ respectively. In a similar fashion, the scaled $\tau_s$ parameters are defined as $(P^4)^{0.25}/\chi$ in the case of fourth rank gradient coefficient; $(P^6)^{0.167}/\chi$ in the case of sixth rank gradient coefficient; $(Q)^{0.25}/\chi$ in the case of fourth rank curvature coefficient; and, $(R)^{0.167}/\chi$ in the case of sixth rank aberration coefficient. In Fig.~\ref{Scaling}, we show the plots of the scaled interfacial energies and the scaled interfacial widths as a function of the scaled tensor coefficients. As expected, as $\tau_s \approx 0$, $\sigma_s$ and $w_s$ reach the equilibrium values calculated for the classical Cahn-Hilliard case, namely, $\sigma  = 0.33$ and $w =4$ for the parameter values chosen by us. However in the case of $\tau_s\gg P^2$, $\sigma_s$ and $w_s$ increase linearly with $\tau_s$. This scaling is very important in choosing appropriate magnitude of tensor coefficients; if they are not of the right magnitude, as noted by ECHTL, the higher order terms have little effect on the interfacial energy anisotropy~\cite{Lowengrub}. The slopes of the linear regime calculated by us (indicated in the figure caption of Fig.~\ref{Scaling}) are, thus, very useful.

\begin{figure}[h]
\centering
\includegraphics[height=2.0in,width=2.0in]{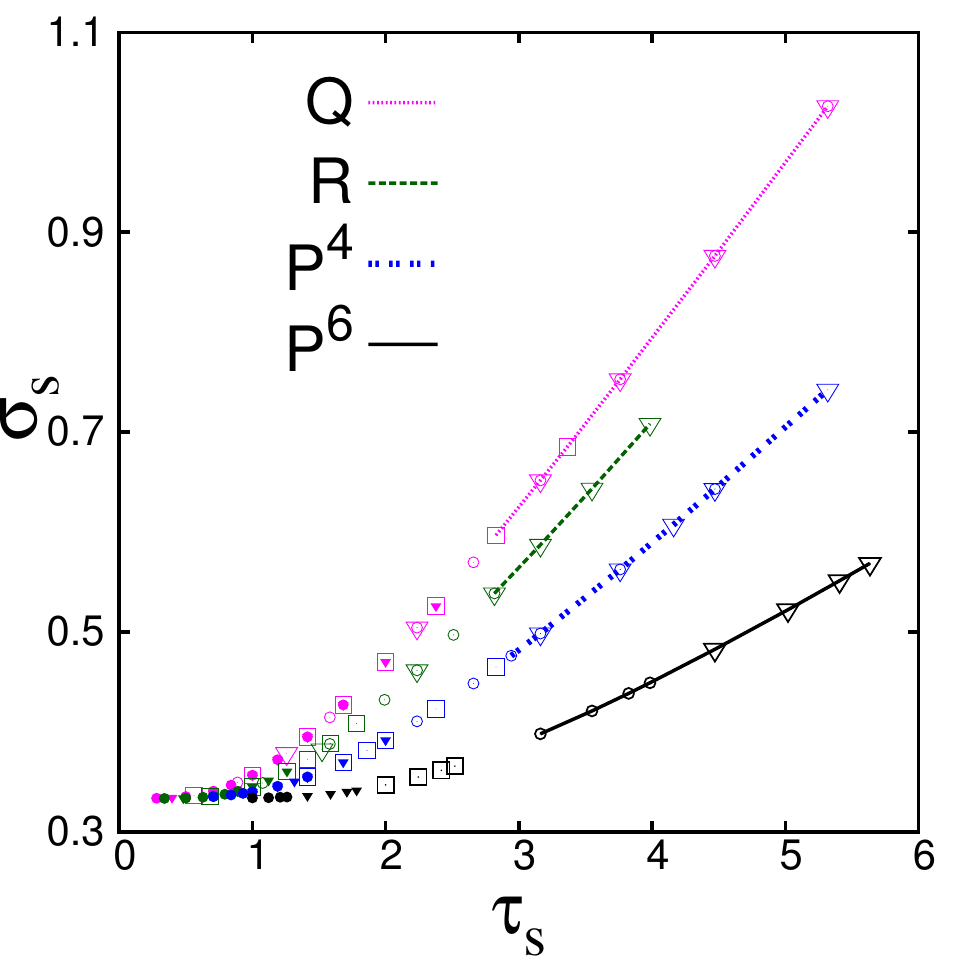}
\includegraphics[height=2.0in,width=2.0in]{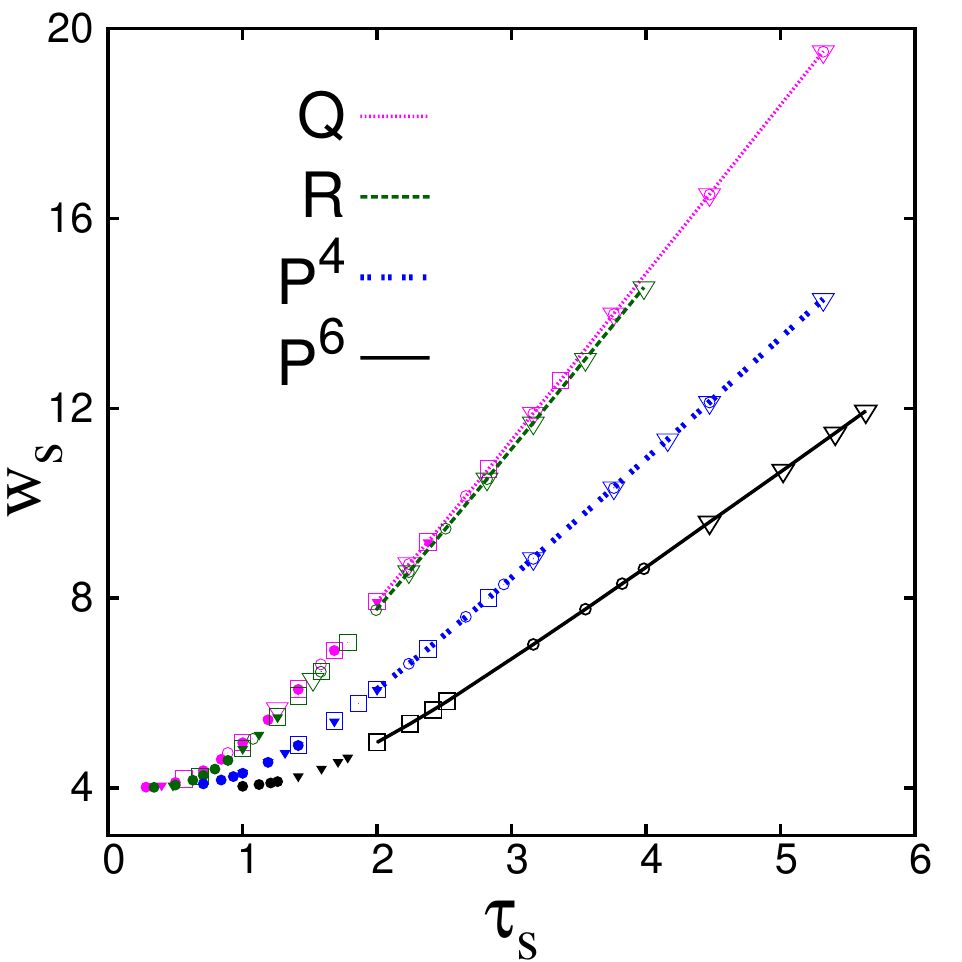}
\caption{ Variation of scaled interfacial free energy $(\sigma_s)$ and width $(w_s)$ with scaled tensor coefficients $(\tau_s)$. Different symbols represent the different values of $P_2$ used in the simulations. Four different line types ( and colors for online-version ) are used to distinguish the linear variation of $\sigma_s$ and $w_s$ with $\tau_s$. The slopes of the linear portion of the scaled interfacial energy ($m_{\sigma}$) are as follows: $Q: 0.173$; $R: 0.145$; $P^4: 0.112$ and $P^6: 0.070$. The slopes of the linear portion of the scaled interfacial width ($m_{w}$) are as follows: $Q: 3.502$; $R: 3.419$; $P^4: 2.485$ and $P^6: 1.935$. Symbols represent different $P^2$ values -- $\triangledown : 0.2$; \Large{$\circ$}\normalsize{$ : 0.4$; $\square : 1$; $\blacktriangledown : 2$ and }\Large{$ \bullet$}\normalsize{$ : 4$} }\label{F:scale}
\end{figure} \label{Scaling}

\subsection{Wulff plots for systems with cubic and hexagonal symmetry}

\begin{figure}[htpb]
\centering
\includegraphics[height=2.0in,width=2.0in]{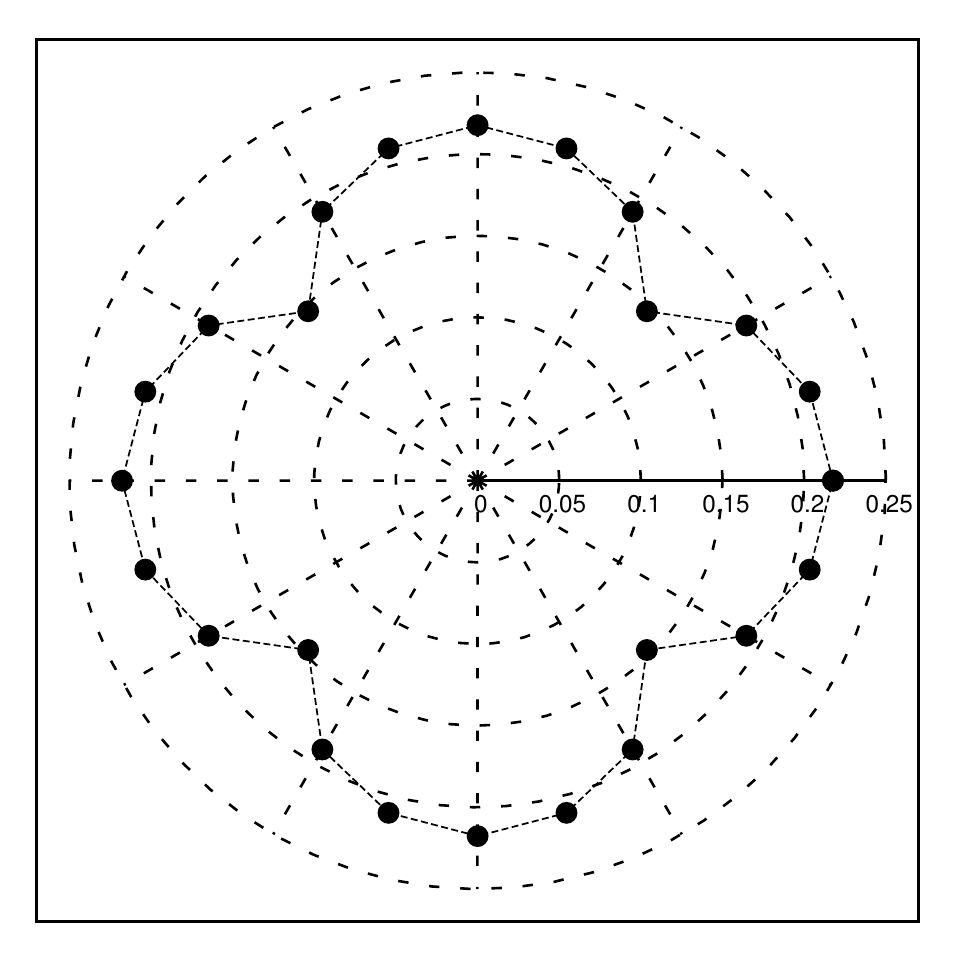}
\includegraphics[height=2.0in,width=2.0in]{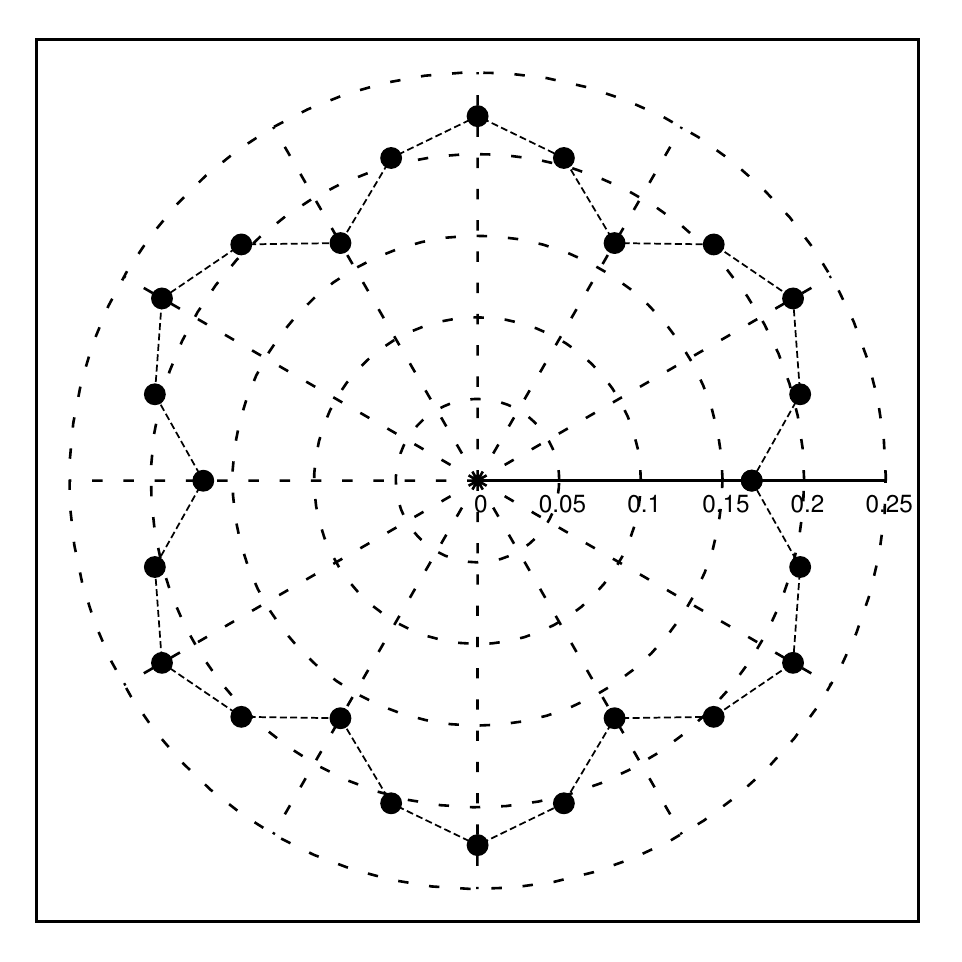}
\caption{ The $xy$-sections of the Wulff plot obtained using $P^6$ for cubic (left) and hexagonal (right) anisotropic systems.}\label{F:wulff}
\end{figure}

Once 1-D profiles can be generated and interfacial energies are calculated, the same calculations can be extended to obtain the variation of interfacial energy with interface orientation; such data on the change of interfacial energy with orientation is typically shown as Wulff plots~\cite{PorterEasterling}. As an example, we show results from a set of calculations in which only $P^6$ tensor coefficient was assumed to be non-zero. We have generated the x-y plane section of the Wulff plots for systems that show (a) cubic symmetry (specifically, one in which the $ \langle 110 \rangle$ directions are preferred over both $\langle 100 \rangle$ and $\langle 111 \rangle$) (left-figure in Fig.~\ref{F:wulff}) and (b) hexagonal symmetry  (right-figure in Fig.~\ref{F:wulff}). For (a), we have used $n_1 = 100$, and $n_6 = -399.9$ and for (b) we have used  $n_1 = 256$ and $n_3 = -210$. Similar Wulff plot sections for other planes and for other systems are possible. However, for the sake of brevity, we only show these two in this paper.  

\subsection{Choice of parameters} \label{ChoiceOfParameters}

From the discussion so far, it is clear that any tensor term can be incorporated as needed -- some choices lend themselves to better numerical implementation while some other lend to better interface profiles. Once the tensor(s) is (are) chosen, the appropriate magnitude of tensor parameters can be identified using the scaling relationship shown in Fig.~\ref{Scaling}. The magnitude of anisotropy, on the other hand, can be tuned using the same simulations which are used to obtain the Wulff plots. Specifically, for choosing parameters which give rise to a specific anisotropy, there are two possibilities -- either we know the interfacial energy or the equilbrium morphology. In both cases, using 1-D simulations, the parameter values can be determined as follows:

\begin{itemize}
\item
In our simulations, using 1D calculations, we can calculate the interfacial energies for different interfaces. Then, as Qin and Bhadeshia~\cite{QinBhadeshia1,QinBhadeshia2} have done, it is possible to tune our parameters in such a way that the ratio of 
interfacial energies of different orientations are the same as that obtained using EAM. 

\item
On the other hand, for solid-solid phase transformations, more complicated equilibrium shapes are possible. Then, the freeware Wulffman~\cite{Wulffman} can be used to identify the relative energies of the facets that lead to the given equilibrium shapes. Again using our 1D simulations and tuning the parameters, we can obtain the requires ratios of interfacial energies for different facets which will result in corresponding anisotropies (and hence equilibrium shapes if the simulations are run to equilibrium).

\end{itemize}

At this point, we wish to note that unlike Qin and Bhadeshia~\cite{QinBhadeshia1,QinBhadeshia2}, we do not get non-equilibrium (concave) boundaries in our simulations. As noted above, the parameters chosen by 
Qin and Bhadeshia include for the cubic case tensors of rank 6 and 8. Hence, no direct comparison is possible. However, it is possible that the method of incorporation of anisotropy, for certain anisotropy values leads to interface-limited kinetics leading to such shapes (as noted in a different setting by Choi et al~\cite{ChoiEtAl2015}). 

Note that there are only two independent components for cubic anisotropy incorporated using P$^4$ or $Q$; on the other hand, using $P^6$ for cubic anisotropy gives three independent components. This is the reason why we are able to get an anisotropy (namely preference of $(110)$ over both $(100)$ and $(111)$) in the sixth rank case. In a similar fashion, we see that there are 5 independent constants in $P^6$ while there are 9 independent constants in $R$. Thus, using $R$, it would be possible to generate more morphologies (that are consistent with hexagonal symmetry) than with $P^6$. Otherwise, we have to incorporate higher order terms (say, for example, 8th rank tensor term). From this, it is clear why Qin and Bhadeshia and Haxhimali et al had to use terms that correspond to 6th and 8th rank tensors to obtain morphologies consistent with cubic anisotropy; even though, they have not made this connection, our way of deriving the polynomials bring this out very nicely.

\subsection{Precipitate morphologies: 2-D simulations}

In this section, we show the cubic (in x-y plane) and hexagonal (in the basal plane) morphologies of precipitates obtained
using $P^4$ and $P^6$ coefficients; it is possible to generate similar microstructures using $Q$ (as also shown by ECHAH) and $R$. However, in this paper, for the sake of brevity, we restrict ourselves to the gradient fourth and sixth rank tensors.

\begin{figure}[htpb]
\centering
\includegraphics[trim=1.6cm 1.6cm 1.6cm 1.6cm,width=0.3\textwidth]{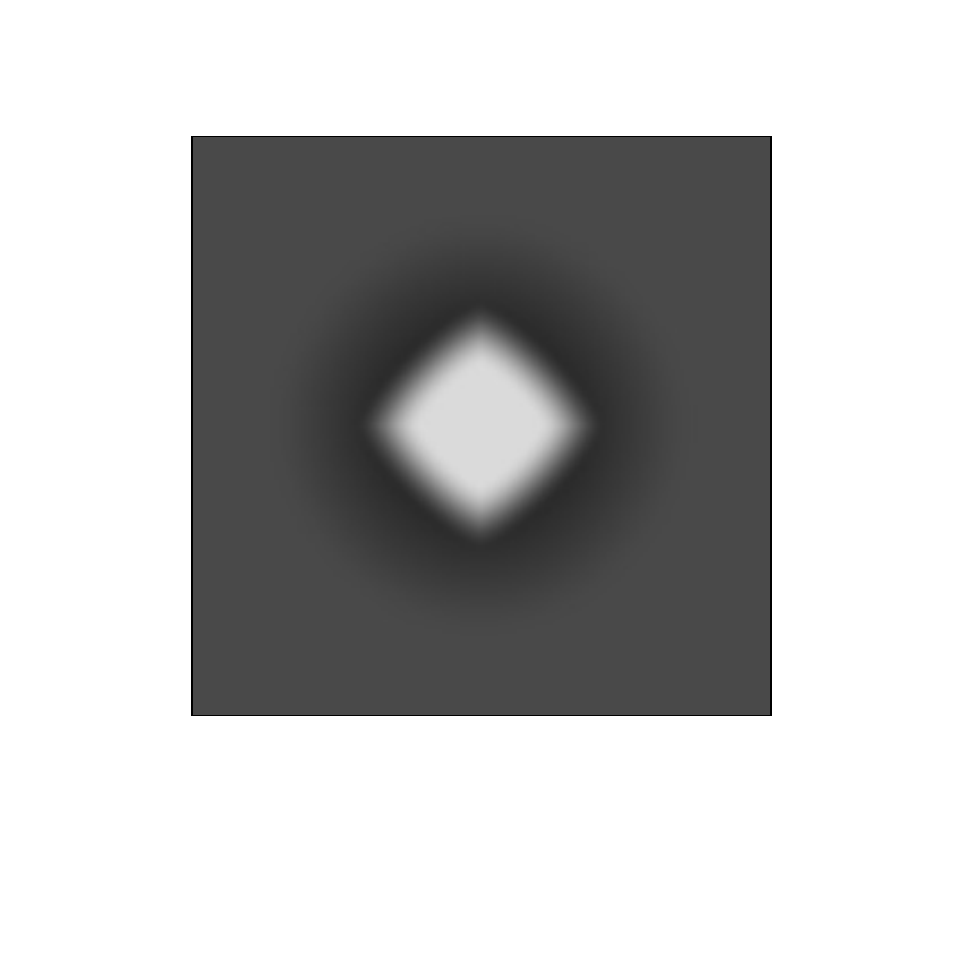}
\includegraphics[trim=1.6cm 1.6cm 1.6cm 1.6cm,width=0.3\textwidth]{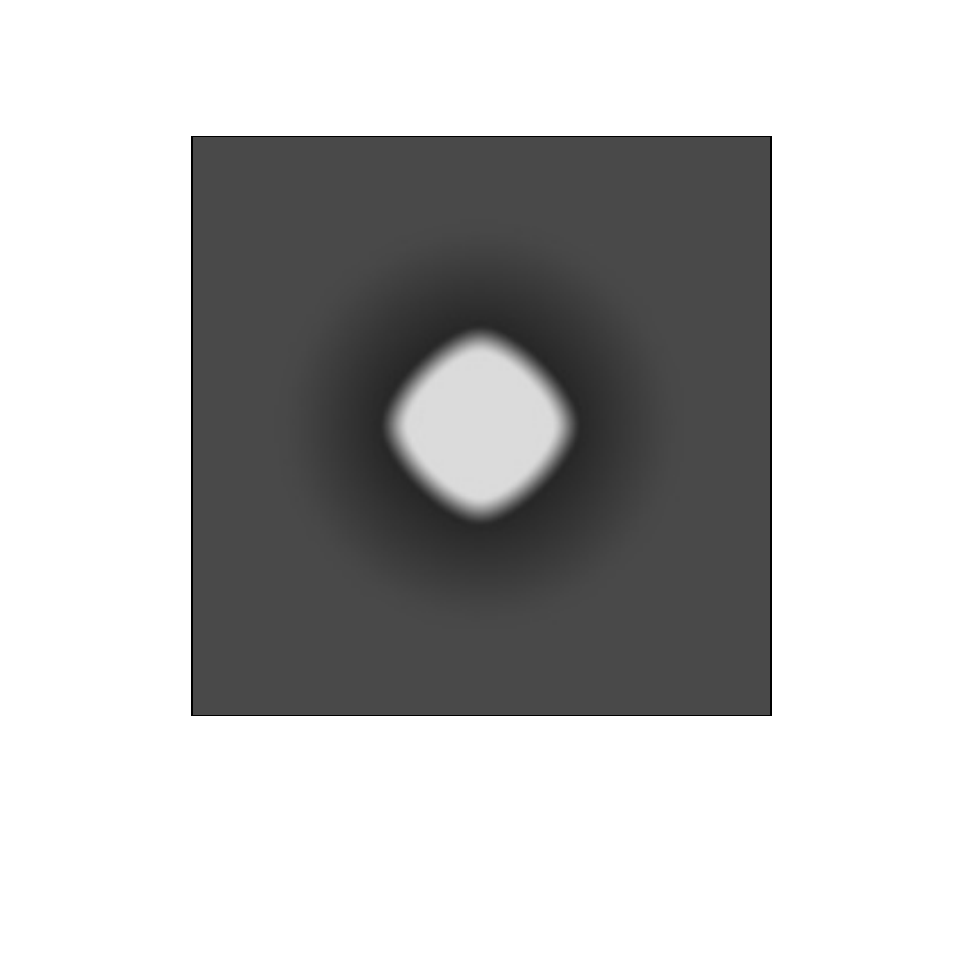} \\
\includegraphics[trim=1.6cm 1.6cm 1.6cm 1.6cm,width=0.3\textwidth]{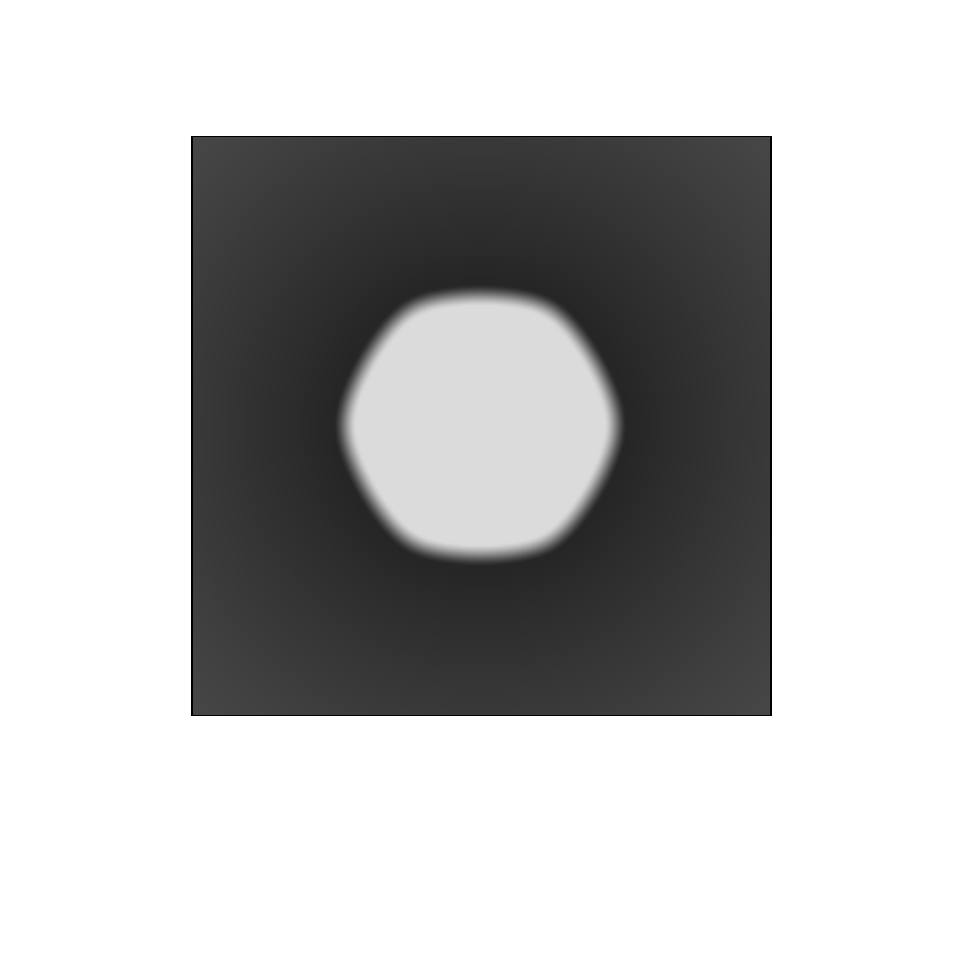}
\includegraphics[trim=1.6cm 1.6cm 1.6cm 1.6cm,width=0.3\textwidth]{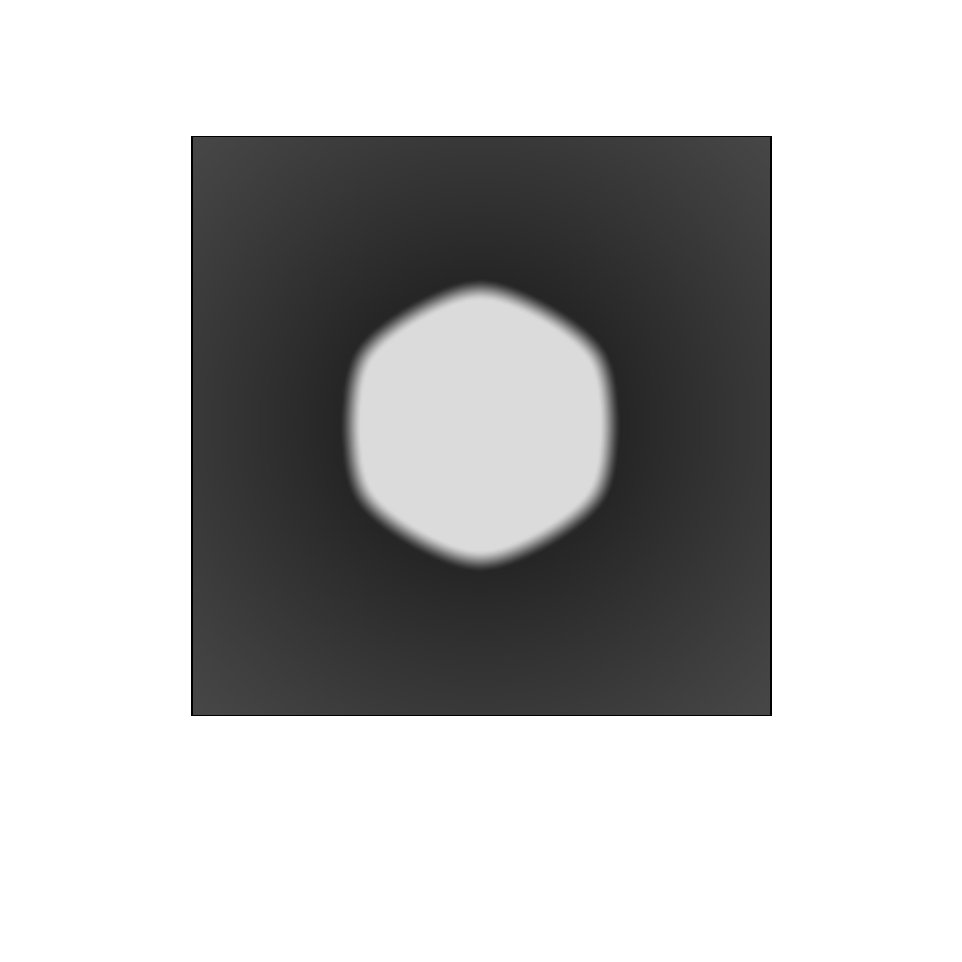}
\caption{ Precipitate morphology with gradient fourth rank (top-left) and sixth rank (top-right) coefficient tensors in system with cubic symmetry. Two figures at the bottom represent morphologies with gradient sixth rank coefficient tensor in system with hexagonal symmetry; Case A : corner along x-axis (bottom-left) and Case B : facet along x-axis (bottom-right) .}\label{2DMorphologies}
\end{figure}

All the 2-D simulations are started with a circular precipitate of radius $10$ units in a system of 256 $\times$ 256 length units with a far field composition of 0.2. In Fig.~\ref{2DMorphologies}, we show the morphologies of precipitates in four different cases -- the first two at the top give rise to cubic morphologies and the next two at the bottom, hexagonal. The top-left figure is the microstructure after 175 time units from a simulation in which only $P^4$ was assumed to be non-zero with $m_1 = 100$ and $m_2 = 320$ . The top-right figure is the microstructure after 175 time units from a simulation in which only $P^6$ was assumed to be non-zero with $n_1 = 100$ and $n_6 = 320$. The parameters in both these cases is chosen such that the $\langle 11 \rangle$ directions are directions of lower interfacial free energy; hence, one can see that the precipitate develops facets in this direction.  On the other hand, the bottom two figures are obtained from simulations (after 1500 time units) in which $P^6$ is assumed to be non-zero; specifically, for the bottom-left figure, we used $n_1 = 32$ and $n_3 = 5$ ( Case A ); and for the bottom-right figure, we used $n_1 = 32$ and $n_3 = -5$ (Case B). In these two cases, again, the parameters are chosen such that the corners (in the bottom-left figure) and the facets (in the bottom-right figure) of the hexagonal precipitate shape are aligned along the x-axis of the simulations. Thus, one can see that using any of the higher order terms, by appropriate choice of the parameters, one can obtain the required morphologies for the precipitates.  

\begin{figure}[htpb]
\centering
\includegraphics[width=.36\textwidth]{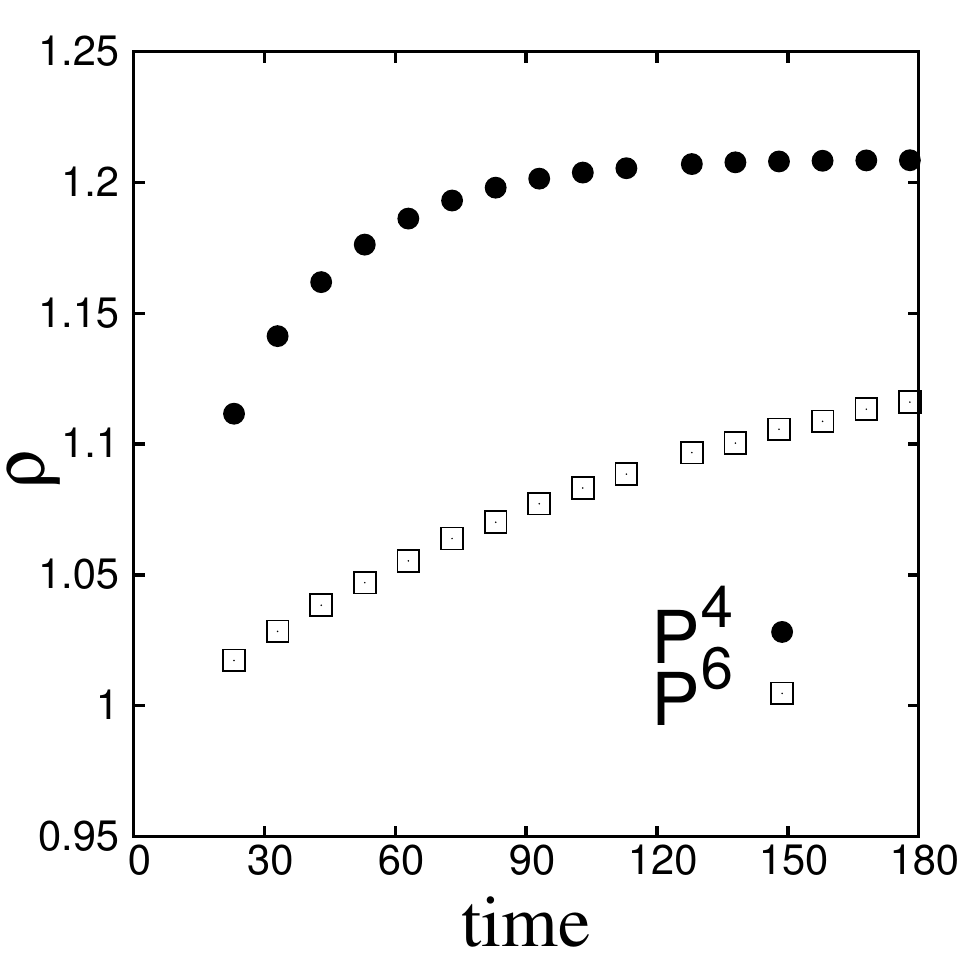}
\includegraphics[width=.36\textwidth]{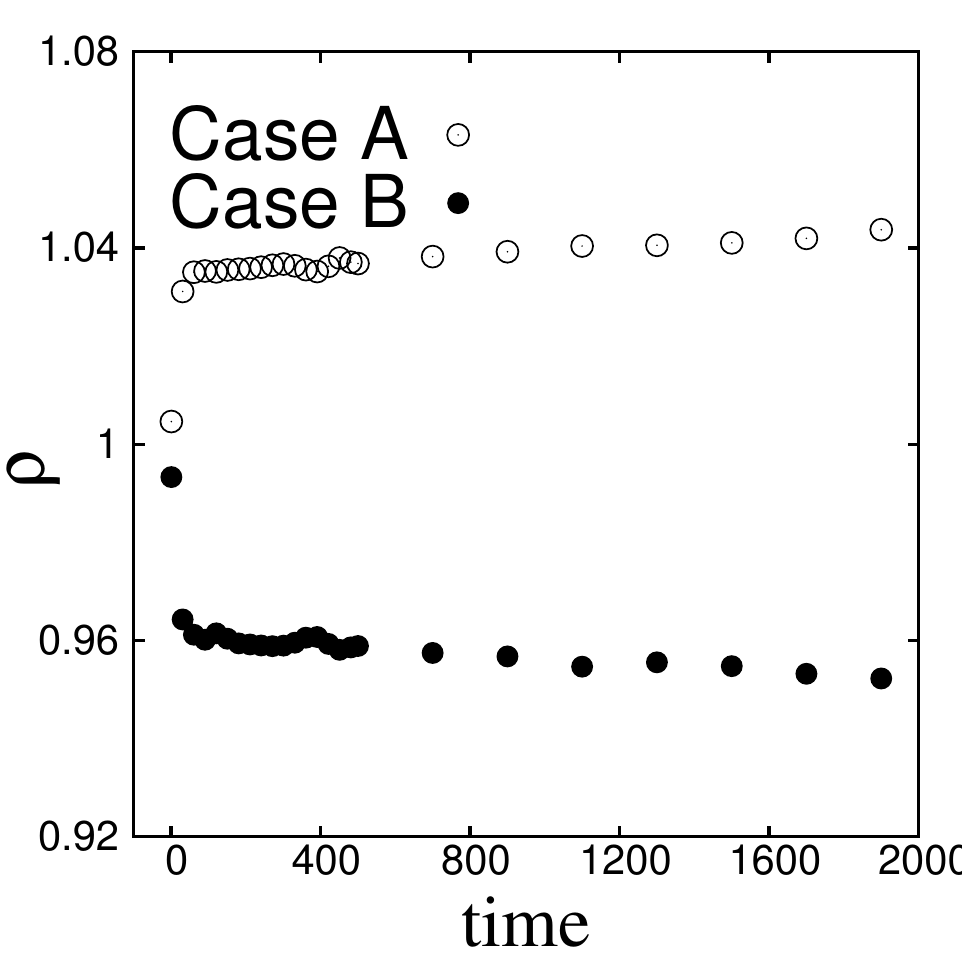}

\caption{ Aspect ratio ($\rho$) of growing precipitate for fourth-rank, $P^4$ and sixth-rank, $P^6$ gradient coefficient tensor in systems with cubic (left) and hexagonal (right) symmetry.}\label{F:aspect}
\end{figure}

In order to understand the precipitate morphologies better, in Fig.~\ref{F:aspect}, we show the aspect ratio $\rho$ of the precipitates as a function of time for cubic (left-figure) and hexagonal (right-figure) precipitates shown in Fig.~\ref{2DMorphologies}. In the case of cubic morphology, the aspect ratio is defined as the ratio of the length of the precipitate along $\langle 10 \rangle$ direction to that along $\langle 11 \rangle$; in the hexagonal case, the aspect ratio is defined as the ratio of the length of the precipitate along $\langle 10 \rangle$ to that along $\langle 01 \rangle$. As one can see, at early stages the aspect ratio is unity (since, the initial shape is a circle). As time proceeds, the aspect ratios change and become less than unity in the cubic case and becomes greater than or less than unity in the hexagonal case depending on the orientation of the hexagon in the 2-D plane and remain more or less a constant during the late stages. In a perfect square, the aspect ratio is expected to be 1.414 ($=\sqrt{2}$) and in a perfect hexagon, the ratio is expected to be 1.16 or 0.87 ($=\cos(\pi/6)$ or $=(cos(\pi/6))^{-1}$) depending on whether the corner or the facet lies along the x-axis. However, the values that we obtain from the simulations are higher / lower than this value; we believe this is because of the slight curvature of the facet; in fact, in the cubic anisotropy case, the shape with a sharper facet has an aspect ratio that is relatively closer to the ideal shape.

\subsection{Cubic symmetry: Dodecahedron in 3-D}

\begin{figure}[htpb]
\centering
\includegraphics[height=1.5in,width=1.5in]{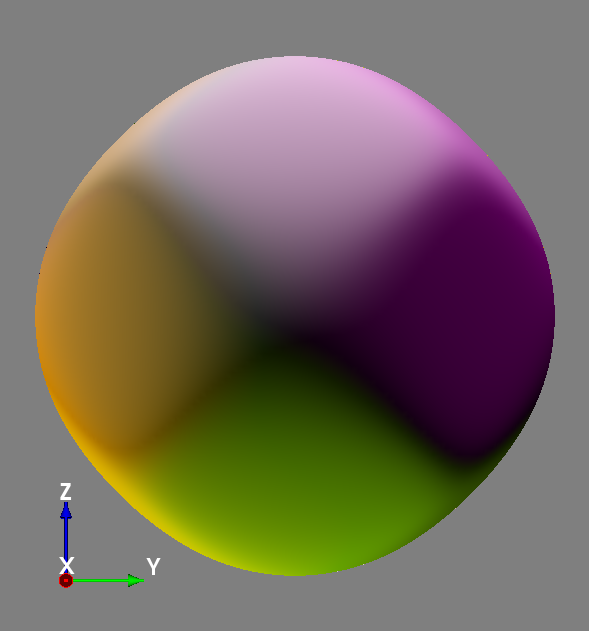}
\includegraphics[height=1.5in,width=1.5in]{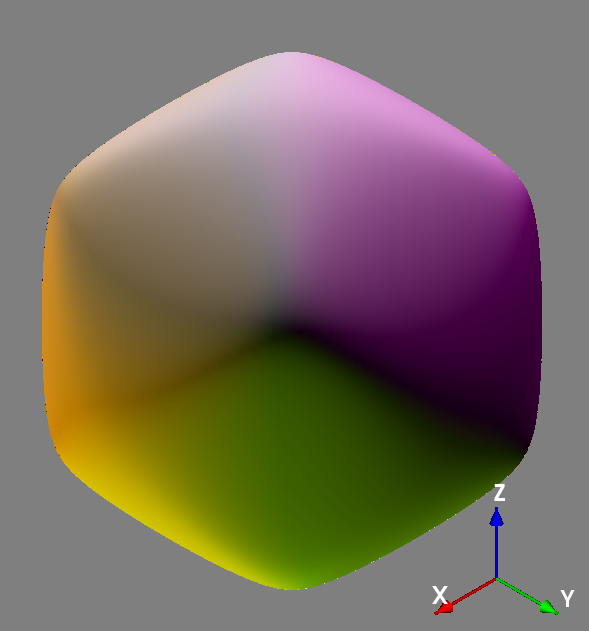}
\caption{View from $\langle001\rangle$ (left) and $\langle111\rangle$ (right) direction of a cubic precipitate morphology with twelve-facet dodecahedron formed by incorporating sixth-rank gradient tensor coefficient.}\label{F:3D_grad_6_dodeca}
\end{figure}

Using the fourth rank tensor terms (either $P^4$ or $Q$ or both), it is possible to obtain cuboidal or octahedral shapes -- that is, shapes which prefer $(100)$ facets or $(111)$ facets~\cite{Roy2015}. However, if the system prefers $(110)$ facets over both $(100)$
and $(111)$, then, we have to necessarily use sixth rank tensors to obtain the equilibrium morphology (namely, dodecahedron: a shape with twelve facets) -- which contains a term of the type $c_1^2c_2^2c_3^2$ which is missing in the fourth rank tensor. In this simulation $P^2$ and $P^4$ are assumed to be isotropic; $P^2$ with $p_1 = 1$ and $P^4$ with $m_1 = 0.001$. The anisotropy is incorporated in the simulation by $P^6$ with $n_1 = 100$, $n_6 = -399.9$ and $n_7 = 5000$. We have grown an initially spherical precipitate of size $10$ units for 5000 time units with a far-field composition of 0.2. The precipitate does become a dodecahedron as shown in Fig.~\ref{F:3D_grad_6_dodeca}. In left figure we show the shape from the side (looking down the x-axis) and in right figure we show the shape as seen from the body diagonal direction (that is, $\langle 111 \rangle$ direction).  

\subsubsection{Hexagonal symmetry: 3-D morphologies of precipitates}

In the case of hexagonal symmetry, several morphologies are possible: for example, hexagonal dipyramid and hexagonal prisms (needles and plates). In the case of hexagonal dipyramids, the facets are along $(10\bar{1}1)$; on the other hand, in the case of hexagonal prism, the facets are along the $(10\bar{1}0)$; depending on the length of the precipitate along the c-axis, these prisms can be needle-like or plate-like. It is also possible to generate equilibrium morphologies which consists of  basal, prismatic and pyramidal facets. Again, it is possible to generate all these morphologies by using either $P^6$ or $R$ or both. 

\begin{figure}[htpb]
\centering
\includegraphics[height=1.5in,width=1.5in]{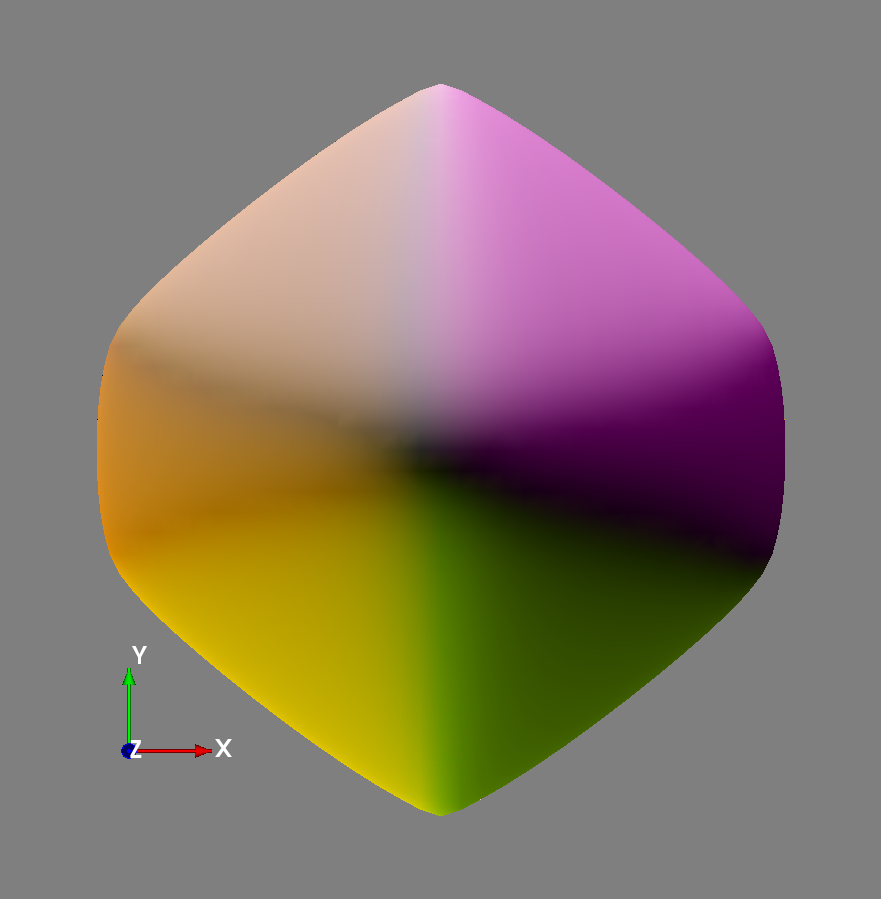}
\includegraphics[height=1.5in,width=1.5in]{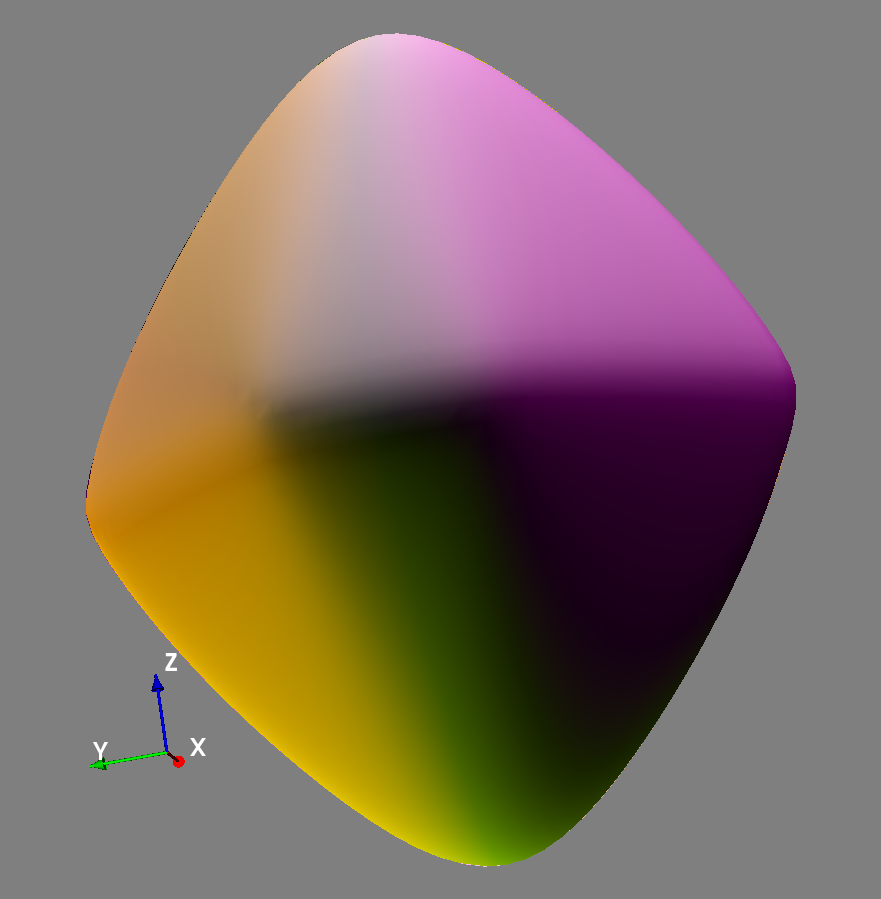} 
\\
\includegraphics[height=1.5in,width=1.5in]{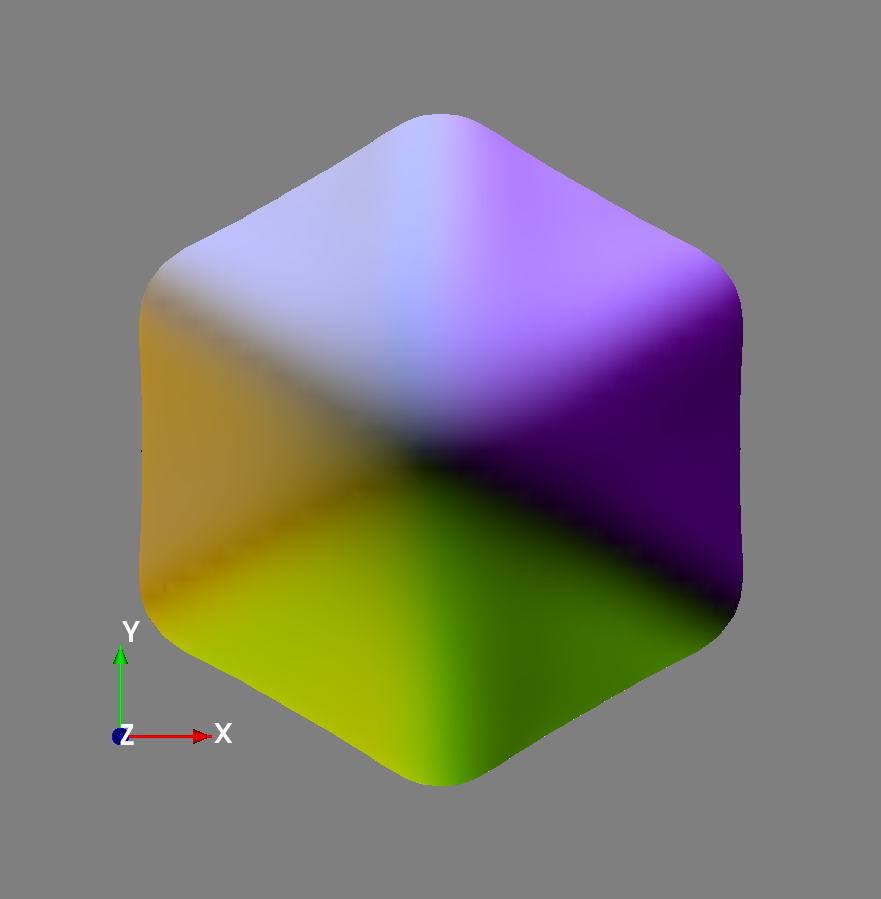}
\includegraphics[height=1.5in,width=1.5in]{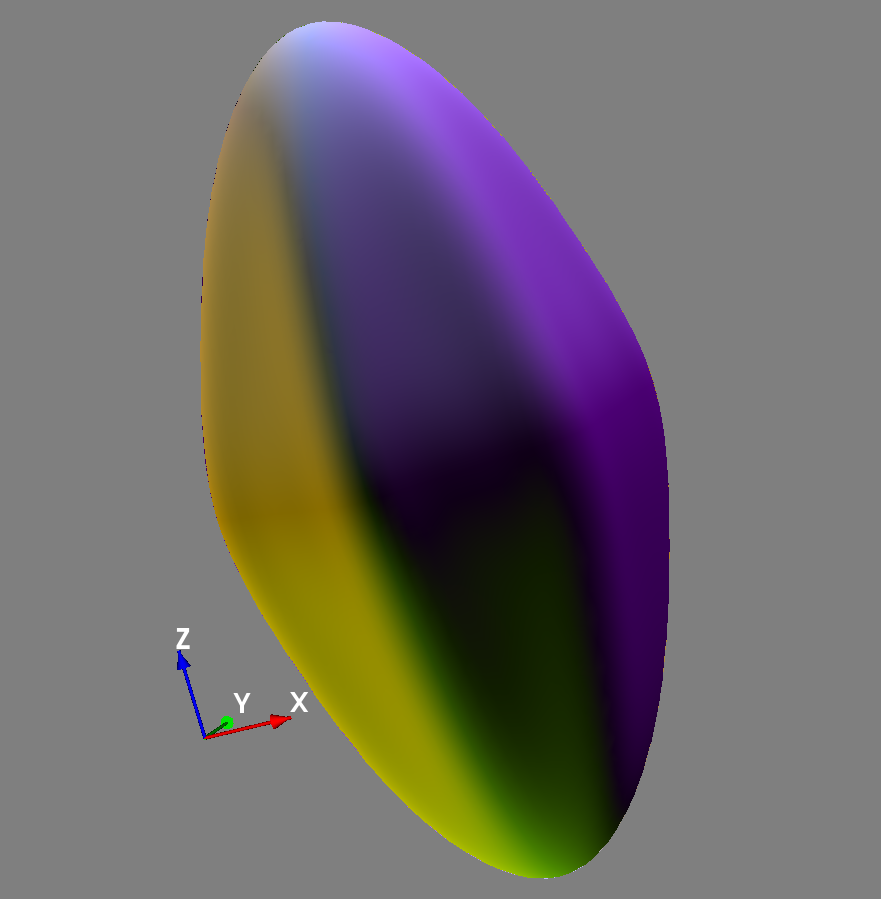}

\caption{Hexagonal dipyramid precipitate morphologies due to the sixth-rank gradient (top) and aberration (bottom) tensor coefficients.}\label{F:HexaDiPyramid}
\end{figure}

In Fig.~\ref{F:HexaDiPyramid}, we show the hexagonal dipyramids obtained using $P^6$ with $n_1 = 100$, $n_2 = 4000$, $n_3 = -9$, $n_4 = 1$ and $n_5 = 1$. (the top-left figure is the view looking down the c-axis and top-right is the view from the side) and using $R$ with $r_1 = 1 $, $r_2 = 600$, $r_3 = 100$, $r_4 = 9$, $r_5 = 1$, $r_6 = -10$, $r_7 = -10$, $r_8 = 1$, $r_9 = 600$ and $r_{10} = -120$ ( the bottom-left figure is the view looking down the c-axis and bottom-right is the view from the side). In both cases, an initial spherical precipitate of size 20 in a matrix with far-field composition of 0.2 was taken. The precipitates were grown for 1000 time units for top figures, and 4000 time units for bottom figures). The morphology in bottom figures have more rounded corners; in addition, the shadow effect seen in the morphology is due to the ripples that come about in the composition profile in the case of aberration terms and the facets themselves are not concave. 

\begin{figure}[htpb]
\centering
\includegraphics[height=1.5in,width=1.5in]{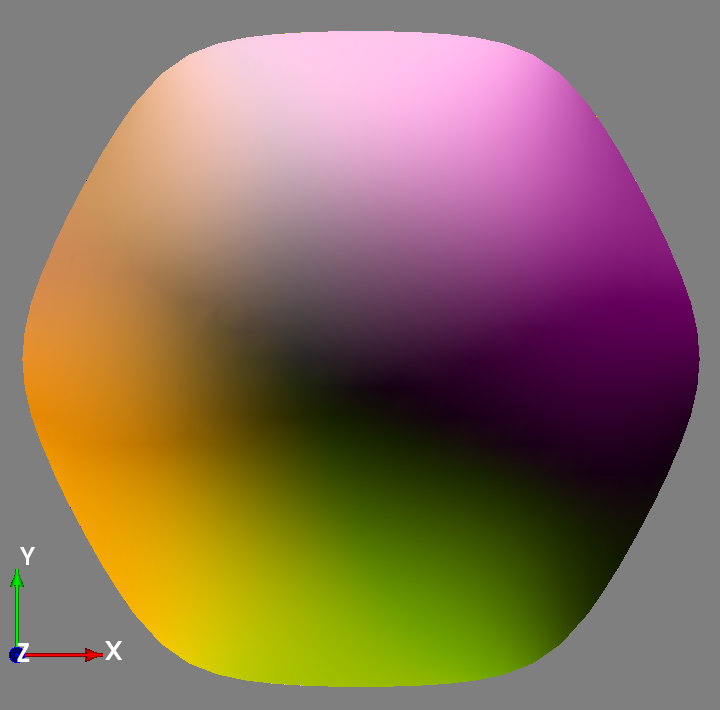}
\includegraphics[height=1.5in,width=0.9in]{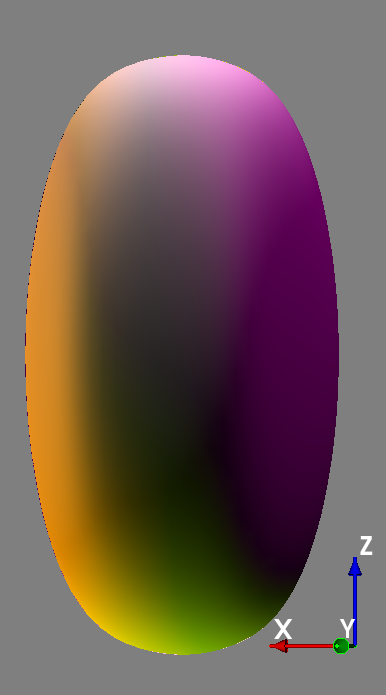}
\caption{Hexagonal prism shaped precipitate due to the sixth-rank aberration tensor coefficient.}\label{F:3D_abb_h_proper}
\end{figure}

In Fig.~\ref{F:3D_abb_h_proper}, we show the morphology in a system with non-zero $R$ with $r_1 = 1 $, $r_2 = 5000$, $r_3 = 1000$, $r_4 = 9$, $r_5 = 3$, $r_6 = 100$, $r_7 = 100$, $r_8 = 1$, $r_9 = 5000$ and $r_{10} = -100$. In this case also, an initial spherical precipitate of size 10 was taken in a matrix with far-field composition of 0.2 and the system was evolved for 500 time units. The precipitate morphology after 100 time units is shown -- with the view down the c-axis as the figure in the left and the view from the side as the figure in the right. We have also seen that by varying the $r$ parameters, it is possible to obtain a hexagonal prism shaped plate instead of a needle as shown here. For the sake of brevity, we do not show such a plate and other morphologies that we obtained in this paper.


\section{Conclusions} \label{Section5}

\begin{enumerate}

\item We have given the free energy (by restricting it to include upto sixth rank tensor terms)
in polynomial form assuming that it depends only on the (coarse-grained)
local composition, its gradient, curvature and aberration. This polynomial form involves several constants; 
we have made a complete listing of the total number of independent constants 
and the constraints on the independent constants (which are different for different symmetry of the 
underlying continua);  

\item We have given a family of extended Cahn-Hilliard evolution equations 
(corresponding to the given free energy functionals);
we have implemented these equations using Fourier spectral and finite difference techniques;
 
\item Using 1-D simulation results, we have characterised the interfaces in terms of their energy, 
anisotropy, shape and width; we have also shown as to how the 1-D simulation results can be used to 
determine the independent constants in any given system knowing either the equilibrium shape of precipitates 
(say, from experiments) or from interfacial energy anisotropy (say from EAM or ab initio calculations); and,

\item We show that it is possible to obtain a wide variety of morphologies in our simulations; 
specifically, in 2-D, we show square and hexagonal precipitate morphologies; in 3-D, we show 
dodecahedron (cubic symmetry) and hexagonal dipyramid, and hexagonal prisms (hexagonal symmetry). Such hexagonal 
faceted precipitate morphologies of precipitates is obtained for the first time using phase field models and 
are relevant for the study of precipitates in many experimental systems.

\end{enumerate}

\section*{Acknowledgements}

We thank T A Abinandanan for useful discussions; Industrial Research and Consultancy Centre, IIT Bombay for financial support (09IRCC16); 
and, PARAM-YUVA at CDAC, Pune, Nebula, Dendrite (DST-FIST facility), and Spinode for computational facilities. 
One of us (AR) thank DST for partial financial support (14DST017).

\bibliography{ArijitGuru}

\end{document}